\documentclass[journal]{IEEEtran}

\usepackage{array,multirow,makecell}
\usepackage[T1]{fontenc}
\usepackage{multirow}
\usepackage{cite}
\usepackage[utf8x]{inputenc} 
\usepackage{longtable}
\usepackage{xcolor}
\usepackage[linesnumbered,ruled,vlined]{algorithm2e}
\usepackage{colortbl}
\usepackage{pgfplots} 

\SetCommentSty{mycommfont}
\usetikzlibrary{patterns}
\usepackage{tikz}
\usepackage{subcaption}
\usepackage{xcolor}
\usetikzlibrary{arrows,shapes,positioning,shadows,trees}
\usepackage{rotating, adjustbox}
\usepackage{pgfplots}
\usepackage{amsmath}
\usepackage{amsfonts}
\usepackage{amssymb}

\ifCLASSINFOpdf
\else
\fi
\usepackage{algorithmic}
\algsetup{linenosize=\small}
\begin{document}
%
\title{Generative AI for Cyber Threat-Hunting in 6G-enabled IoT Networks}

\author{
\IEEEauthorblockN{Mohamed~Amine~Ferrag\IEEEauthorrefmark{1}, Merouane Debbah\IEEEauthorrefmark{1}, Muna Al-Hawawreh\IEEEauthorrefmark{2}}\\
 \IEEEauthorblockA{\IEEEauthorrefmark{1}Technology Innovation Institute, 9639 Masdar City, Abu Dhabi, UAE}\\
 \IEEEauthorblockA{\IEEEauthorrefmark{2}School of Information Technology, Deakin University, Australia}\\
 \IEEEauthorblockA{Emails: mohamed.ferrag@tii.ae, merouane.debbah@tii.ae, muna.alhawawreh@deakin.edu.au}
 }
\maketitle

\begin{abstract}
The next generation of cellular technology, 6G, is being developed to enable a wide range of new applications and services for the Internet of Things (IoT). One of 6G's main advantages for IoT applications is its ability to support much higher data rates and bandwidth as well as to support ultra-low latency. However, with this increased connectivity will come to an increased risk of cyber threats, as attackers will be able to exploit the large network of connected devices. Generative Artificial Intelligence (AI) can be used to detect and prevent cyber attacks by continuously learning and adapting to new threats and vulnerabilities. In this paper, we discuss the use of generative AI for cyber threat-hunting (CTH) in 6G-enabled IoT networks. Then, we propose a new generative adversarial network (GAN) and Transformer-based model for CTH in 6G-enabled IoT Networks. The experimental analysis results with a new cyber security dataset demonstrate that the Transformer-based security model for CTH can detect IoT attacks with a high overall accuracy of 95\%. We examine the challenges and opportunities and conclude by highlighting the potential of generative AI in enhancing the security of 6G-enabled IoT networks and call for further research to be conducted in this area.
\end{abstract}

\begin{IEEEkeywords}
Generative AI, Security, GPT, GAN, IoT, 6G.
\end{IEEEkeywords}

%
\IEEEpeerreviewmaketitle

\section{Introduction}
\begin{figure}[t!]
\centering
    \includegraphics[width=0.8\linewidth]{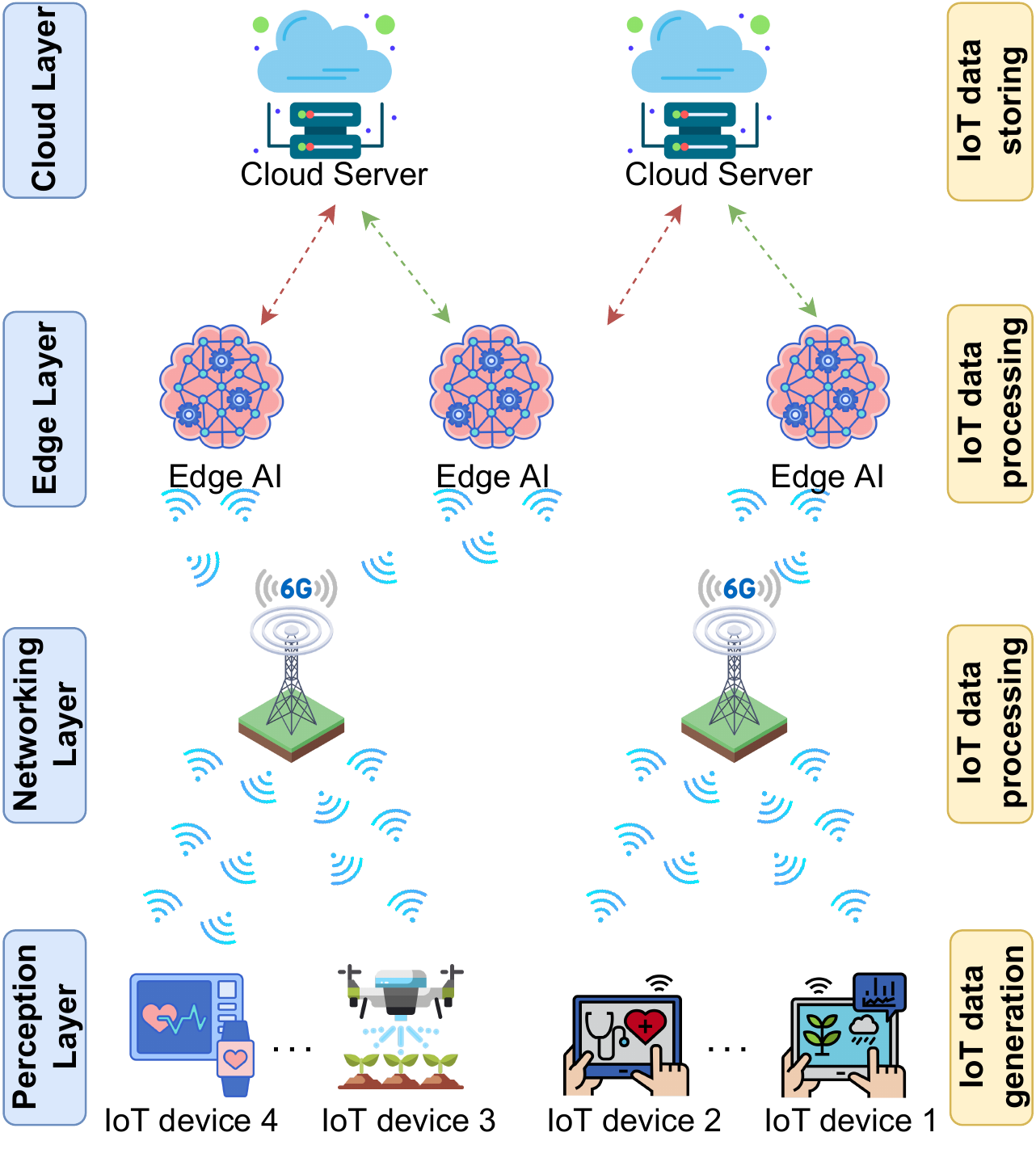}
    \caption{A 6G-enabled IoT Network.}
    \label{fig:fig2}
\end{figure}

The Internet of Things (IoT) has revolutionized how people interact with the environment around them. With the emergence of 6G technology, the IoT is expected to reach new levels of connectivity and intelligence \cite{gerodimos2023iot,babaghayou2021whisper}. As shown in Figure in Fig. \ref{fig:fig2}, the 6G-enabled IoT network comprises four tiers: the perception tier, the network tier, the edge tier, and the cloud tier. The perception layer is the first layer of the 6G-enabled IoT network. The layer is in charge of sensing and collecting data from the physical world. Various sensors, such as temperature, microphones, and camera sensors, are embedded in devices such as smartphones, smart home appliances, and industrial equipment at this layer. The network layer is the second layer of the 6G-enabled IoT network. It is responsible for the connection of all the devices in the network and allowing data transfer between them. The network layer is composed of various networking technologies, such as Wi-Fi, Bluetooth, and 6G, which enable the devices to communicate with each other and with the edge layer. The edge layer is the third layer of the 6G-enabled IoT network. It's responsible for processing and analyzing data at the edge of the network, rather than in the cloud. The edge layer is composed of peripheral devices, such as routers, gateways, and servers, which are equipped with powerful processors and memory. The cloud layer is responsible for the storage, management, and analysis of data collected by the perception layer. This layer consists of cloud servers, located in data centers and accessible via the Internet.

Generative AI refers to a class of artificial intelligence that can generate new material, such as music, images, or text \cite{dhariwal2020jukebox}. Those particular systems are constructed to learn the characteristics and features of a specific dataset and then use that intelligence to generate new, original content that follows the same pattern. Generative AI in general has a variety of different uses, including data synthesis, algorithm Invention, data augmentation, and anomaly detection. Table \ref{tab:tab4} presents the comparison between the Generative AI model and the Traditional AI.

There are, at the same time, limitations on the use of generative AI for cyber threat hunting. One challenge is that such systems depend on the completeness and quality of the data on which they're trained. If the data being trained is incomplete or biased, the performance of the system can be damaged. In addition, generative AI systems can create false positives, which involves the identification of a threat that could be occurring when no threat really exists. This can result in useless testing and resource consumption

Cyber threat hunting is the process of searching for proactive signs of malicious activity within an organization's networks and systems. This is key to finding and minimizing the threats before they cause serious damage. Over the past few years, however, an increasing interest has emerged in the application of generative artificial intelligence (AI) to cyber threat hunting. There are many different ways to use generative AI for cyber threat hunting, and one of them is to analyze large quantities of data to find and identify patterns and anomalies that could provide an indication of the presence of a threat. One method of using generative AI for cyber threat hunting is the report and alert generation. The traditional cyber threat hunt can often include reviewing massive quantities of data manually and then generating alerts or reports depending on the findings. This is time-consuming and can lead to errors. Generative AI systems, on the other hand, can perform real-time data analysis and produce reports or alerts directly depending on the findings. This can contribute to a significant acceleration of the threat detection process and decrease the number of missed security threats. Using generative AI for cyber threat hunting is one technique that is adopted for analyzing large amounts of data in order to find any abnormal activity patterns and anomalies that might reveal the existence of a malicious threat. A generative AI system, for example, might be trained on a network traffic log dataset and then be employed to recognize abnormal patterns of activity that may be indicative of a threat adversary.

Motivated by the facts mentioned above, in this article, we review the use of generative AI for cyber threat-hunting in 6G-enabled IoT networks. Specifically, we discuss the Generative AI use cases for IoT applications as well as evaluate three generative AI models for cyber security, including, the GAN-based method, GPT-based method, and BERT-based method. Therefore, we propose a new GAN and Transformer-based model for Cyber Threat-Hunting in 6G-enabled IoT Networks. The experimental analysis results with a new cyber security dataset demonstrate that the Transformer-based security model for cyber threat-hunting can detect IoT attacks with a high overall accuracy of 95\%. In addition, we provide several challenges regarding the use of generative AI for cyber threat-hunting in 6G-enabled IoT networks, including, scalability issues, decentralized training issues, data quality issues, energy challenges, privacy-preserving challenges, and tokenization challenges.

\begin{table*}[t!]
\centering
\setlength{\tabcolsep}{2.5pt}
\renewcommand{\arraystretch}{1}
\caption{Comparison between Generative AI model and Traditional AI.}
\centering
\scriptsize
\label{tab:tab4}
\begin{tabular}{|p{0.6in}|p{2.5in}|p{2.5in}|}
\hline
\centering
\textbf{Metric}   & \textbf{Generative AI}    & \textbf{Traditional AI}    \\ \hline \hline
 Data Availability & Demand large amounts of data to train, which makes it challenging to use in situations where data is limited & Requires less data to train, which makes them more feasible to use in situations where data is limited.\\ \hline
 Scalability  & Generative AI models can be computationally demanding, which makes them challenging to scale & Traditional ML models are often less computationally demanding, which makes them more scalable\\ \hline
 Data Privacy	 &  Generative AI models usually involve access to a large dataset, which may be owned by a third party  & Traditional AI models are based on user or organization-owned data\\ \hline
 Data security    &  Generative AI models can pose a higher privacy risk since they generate new data that may not be fully controllable & Traditional AI models use existing data and can be less risky regarding data privacy\\ \hline
 Adversarial Robustness & Generative AI models can be vulnerable to adversarial attacks, as these models are trained to generate new data rather than classify the existing data & Traditional AI models are often more resilient to adversarial attacks than generative AI, as they are trained to classify the existing data rather than generate new data\\ \hline
 Overfitting	 & Generative AI models can be susceptible to overfitting & Traditional AI models are often less susceptible to overfitting compared to Generative AI models\\ \hline
Transfer Learning	  & The transfer of generative AI models to different tasks or domains is challenging  & Traditional AI models are often easier to transfer to new tasks or domains\\ \hline
 Bias	 & Generative AI models can be vulnerable to bias & Traditional ML models are often less vulnerable to bias compared to Generative AI models\\ \hline
Time Complexity		 & Generative AI models can be computationally intensive & Traditional AI models are often less computationally intensive compared to Generative AI models\\ \hline
Energy cost		 & Require more computational power and energy  & Require Less computational power and energy compared to generative models\\ \hline
\end{tabular}
\end{table*}

\begin{table*}[t!]
\centering
\setlength{\tabcolsep}{2.5pt}
\renewcommand{\arraystretch}{1}
\caption{Recent works on Generative AI for Cyber Threat-Hunting.}
\centering
\tiny
\label{tab:tab1}
\begin{tabular}{|p{0.5in}|p{0.2in}|p{2.1in}|p{0.5in}|p{0.4in}|p{0.4in}|p{2.5in}|}
\hline
\textbf{Framework}                                          & \textbf{Year} & \textbf{IoT Network   model} & \textbf{Datasets} & \textbf{Gen.   AI} & \textbf{ML model} & \textbf{ Pros (+) Open Issues (-) } \\ \hline
Zhang et al. \cite{zhang2020poisongan}       & 2021          &     By maintaining a singular global model, an IoT network disseminates the model throughout a cloud server and various edge nodes or clients  & MNIST, Fashion-MNIST, and CIFAR10  &  GAN         &   CNN      &   + The proposed generative poisoning method is efficient against a Federated learning framework \newline - Poisoning attack on federated learning may require a lot of computational power   \\ \hline
Ranade et al. \cite{ranade2021cybert}  &   2021 & the paper's design requires the assembly of unlabeled textual data on cybersecurity from a variety of sources such as the National Vulnerability Database, and open source blogs.  & Common Vulnerabilities and Exposures  & BERT    &  MLM   &   + The model that has been fine-tuned can carry out multiple cybersecurity-related tasks with great precision and effectiveness  \newline - The paper does not discuss the scalability of the model and its ability to handle large-scale datasets, which may be a limitation in real-world applications \\ \hline
Wu et al. \cite{wu2021intelligent}           & 2021          &  At the edge nodes of the network, the IDS is positioned to gather historical information about the network's edges. This information is then transferred to a data center where feature selection and model training takes place.            & CIC-DDOS2019 and CIC-DDOS2019      &     GAN  &   CNN      &     + Improves the accuracy of intrusion detection systems  \newline - The proposed method requires a significant amount of computational resources, which may not be feasible for resource-constrained IoT edge nodes              \\ \hline
Cui et al. \cite{cui2021security}            & 2021          &   At the edge nodes of the network, the IDS is positioned to gather historical information about the network's edges. This information is then transferred to a data center where feature selection and model training takes place. & KDD CUP 1999   &   GAN         &     CNN     &    + This framework utilizes blockchain technology to ensure data integrity and prevent single-point failure       \newline - The system is only tested on one specific dataset, which may not generalize to other datasets or real-world situations          \\ \hline
Ranade et al. \cite{ranade2021generating}    & 2021  &         There are three primary sources that make up the CTI collection, which include technical APT reports, vulnerability databases, and security news articles.    &  WebText
dataset  & GPT   &  FNN     &  +  Provides a method for automatically generating fake CTI text using Transformers to perform a data poisoning attack \newline - The paper does not provide a solution for detecting or mitigating the effects of a data poisoning attack using fake CTI              \\ \hline
Yang et al. \cite{yang2022generative}        & 2022          &  The 6G architecture is divided into four tiers, including marine, terrestrial, aerial and satellite networks, and utilizes AI services throughout the process, from data sensing to smart applications  &  Simulation   &  GAN    &     AE    &       + Determine the trustworthiness of network devices and make trust management more resilient \newline The proposed framework does not discuss any potential security risks or vulnerabilities that may arise from using the proposed GAN method   \\ \hline
Tabassum et al.   \cite{tabassum2022fedgan}  & 2022          &   The system focuses on two main components: IoT devices and Edge Nodes &  UNSW-NB15, NSL-KDD&        GAN      &    CNN     &     + The model is distributed over IoT devices and uses augmented local data for training \newline -  Efficient models require ample and varied training data in FL       \\ \hline
Jo et al. \cite{jo2022vulcan}  & 2022          &  The architecture is separated into two sections: the collection of CTI data and the utilization of CTI data. The collection section comprises five elements: a scraper of data, a pre-processor, an identifier of threat entities, a linker of entities, and an identifier of threat relationships &  540 K articles about cyber threats   &   BERT     &  BiLSTM  &     + For both NER and RE tasks, there was an average F-score of 0.972 and 0.985, respectively
 \newline - Vulnerable to adversarial attacks, where attackers can manipulate the input to the model in order to produce a desired output  \\ \hline
Habibi et al.   \cite{habibi2023imbalanced}  & 2023          &    The IoT system is made up of various components including sensors, wireless power, microcontrollers, and antennas which are connected to a gateway. They are linked to the cloud computing   &  Bot-IoT dataset  & GAN      & MLP           &     + The proposed solution has shown promising results, achieving a high accuracy rate of 98.93\%, an F1-score of 0.9907, and geometric-score values of 0.9874  \newline - The proposed solution may not be effective in dealing with zero-day threats  \\ \hline
\end{tabular} \\
The abbreviations BiLSTM, ML, FL, CNN, GAN, GPT, CTI, FNN, AE, MLP, BERT, MLM, NER, and RE stand for A Bidirectional Long Short-Term Memory, Machine Learning, Federated Learning, Convolutional Neural Network, Generative Adversarial Networks, Generative Pre-trained Transformer, Cyber Threat Intelligence, Feedforward Neural Network, Auto Encoder, Multilayer Perceptron, Bidirectional Encoder Representations from Transformers, Masked Language Modeling, Named Entity Recognition, and Relation Extraction, respectively.
\end{table*}

This article is structured as follows: In section \ref{sec:2}, the generative AI use cases for IoT applications are presented. Section \ref{sec:3} outlines the methodology of generative AI for Cyber Security. The proposed GAN and Transformer-based model is detailed in section \ref{sec:4}, while the results of our experiments are provided in section \ref{sec:5}. Section \ref{sec:6} covers the open challenges concerning the use of generative AI for cyber threat-hunting in 6G-enabled IoT networks, and finally, section \ref{sec:7} concludes the article.

\section{Generative AI Use Cases} \label{sec:2}

As GPT and GAN are two completely different models with different strengths and weaknesses, they can be combined to form a more robust system. For example, GPT can be used to generate text-based synthetic data, which can then be transmitted through a GAN to generate realistic images. This combination can be employed to generate more synthetic data for computer vision models, audio models, security models, and text models, which can help improve their robustness and accuracy. In this section, we discuss the Generative AI use cases for IoT applications, including, visual IoT applications, audio IoT applications, text-based IoT applications, code-based IoT applications, and IoT security.

\subsection{Visual IoT Applications}
Visual applications can be employed in various types of IoT contexts, from surveillance and real-time monitoring to diagnostic and remote maintenance. Specifically, as an example, video cameras and other sensors can be embedded in IoT systems to facilitate the monitoring of equipment or installations remotely, enabling users to identify and solve problems quickly before they progress to serious issues. One of the most well-known applications of generative AI is the application of generative adversarial networks (GANs) to generate realistic images. A GAN comprises a pair of neural networks: one is a generator, which learns to produce novel images, and the other is a discriminator, which learns to distinguish genuine from fabricated images. Both networks are trained jointly in a zero-sum game, where the generator tries to generate images that are unrecognized from the real ones and the discriminator tries to categorize the images as real or fake with accuracy. We consider a visual IoT system used to monitor the status of crop health in an agricultural field \cite{friha2021internet}. Specifically, the system could employ cameras and sensors to capture data on the crops' development and progress. Through the use of generative AI, however, it is possible to produce pictures or videos that demonstrate how crops are expected to develop over time, based on the data that has been collected. This could be used to assist in helping farmers to better understand how to properly manage their crops and ensure that they are making the best use of their resources. As such, Generative AI will play a key role in the future development of visual IoT applications. 

\subsection{Audio IoT Applications}
An example of a voice-based device in the IoT is the Amazon Echo, which employs the Alexa voice assistant to enable consumers to monitor smart home equipment and control and access content via their voice prompts. The device can be deployed to power on and off lamps, regulate room thermostats, listen to music, and many others. Additional examples of voice-activated devices in the IoT include Apple HomePod, Google Home, ...etc. The use of voice-activated devices is a one-way audio application that can be implemented in the IoT. Therefore, Generative AI has the power to transform the way we interact with audio applications in 6G-enabled IoT networks. By employing machine learning algorithms to generate new audio content, generative AI can enable a vast number of applications that were not previously possible. One possible potential application of generative AI in audio is the construction of personal audio environments for virtual and augmented reality (VR/AR) experiences. By processing the preferences of a user and the context of the VR/AR environment, a generative AI system can generate a specific audio experience that is designed to immerse the user in the virtual reality environment. While this could be particularly beneficial for entertainment and gaming apps, it could also be implemented in more practical environments, such as training simulations. In addition, Generative AI could also be used to enhance the availability of audio content for people with hearing difficulties. By generating descriptions of visual content, such as films or television programs, generative AI could allow people with hearing loss to access and enjoy audio-visual media that was previously unfeasible.

\subsection{Text-based IoT Applications}
A specific application of Natural Language Processing (NLP) in the IoT is the development of automated intelligent assistants and chatbots. The systems use NLP-based algorithms to interpret and process user queries, enabling people to interact with machines and systems using natural language. To illustrate, a user may ask a smart assistant to activate the lights or regulate the temperature in their residence, and the assistant will employ NLP to interpret and process the query. Therefore, a popular implementation of generative AI involves the deployment of language models to generate text in natural language. These models are trained on larger text datasets and can produce sentences and paragraphs consistently that reflect the style and structure of the language. These models can be employed for various applications, such as dialogue systems, text summarization, and machine translation. Generative AI has also potential uses in augmenting data, where it can be employed to produce more training examples for machine learning models. This may especially be relevant in situations where the quantity of training data available is small, as it can help to enhance the model's performance. However, a major issue of concern with generative AI is the possibility that it can be exploited to generate malicious or fraudulent material, such as deepfake videos or fake news articles. In order to deal with this issue, it's important to build detection and mitigation approaches for such threats, such as enhancing the resilience of discriminative models and building methods for authenticity verification of the produced contents.

\subsection{Code-based IoT Applications}
A key feature of the IoT is the generation of codes, or software programs, that allow devices to interact with each other and execute specific actions. These codes can be employed to build applications, which can be downloaded to various devices and used to monitor and operate various components of the IoT. There are various approaches to employing generative AI for code generation. There is one approach that involves the use of the machine learning model to generate code based on a set of required input parameters and preferences. Specifically, for example, a user could provide the intended function of a code fragment and the machine learning model will generate the code that will accomplish the desired function. This might be especially valuable in situations where the intended function is complicated and takes time for a human developer to code manually. However, there are also potential limitations and challenges when using generative AI for code generation. One challenge is that the training data for the machine learning model needs to be high quality in order to have the model generate reliable code. An additional challenge is that generated code is not always able to be interpreted by humans, which may make it challenging for the developers to understand and debug.

\subsection{IoT Security}
GPT and GAN are two popular machine learning models that have been widely used in various applications such as natural language processing (NLP) and image generation. Although both models are generative, they have different functions and applications in the security field. With the increase of connected devices, security has now become a major issue. These devices can be susceptible to attackers, who can use them to gain access to sensitive data or to monitor the devices themselves. One potential use of generative AI in IoT security is in detecting and preventing cyberattacks. Most cyberattacks are actually automated and employ pre-determined attack patterns to attempt to penetrate systems. Generative AI can be employed to recognize and prevent these patterns, essentially stopping the attack before it causes any serious damage. Generative AI can also be used to monitor devices continuously for abnormal patterns or activities, allowing for quick recognition and resolving potential security risks. Another potential use of generative AI in IoT security is the creation of secure communication protocols. Multiple IoT devices depend on wireless communication to communicate with the Internet and with each other. These wireless communications, however, can be easily captured and compromised by attackers. Generative AI can be employed to build secure communication protocols that encrypt data transmitted between devices, which makes it significantly more difficult for attackers to gain access to critical data. 

\begin{table}[t!]
\centering
\setlength{\tabcolsep}{2.5pt}
\renewcommand{\arraystretch}{1}
\caption{Comparison between Generative AI models.}
\centering
\scriptsize
\label{tab:tab2}
\begin{tabular}{|p{1.6in}|p{0.5in}|p{0.5in}|p{0.5in}|}
\hline
\centering
\textbf{Metric}                    & \textbf{GAN}                                            & \textbf{GPT}               & \textbf{BERT}                                \\ \hline \hline
Model type                         & Generative                                              & Transformer          & Transformer                            \\ \hline
Anomaly detection                  & Yes                                                     & Yes                        & Yes                                          \\ \hline
Authentication                     & No                                                      & No                         & No                                           \\ \hline
Encryption and decryption          & No                                                      & No                         & No                                           \\ \hline
Network intrusion detection        & Yes                                                     & Yes                        & Yes                                          \\ \hline
Access control                     & Yes                                                     & Yes                        & Yes                                          \\ \hline
Phishing detection                 & Yes                                                     & Yes                        & Yes                                          \\ \hline
Spam detection                     & Yes                                                     & Yes                        & Yes                                          \\ \hline
Malware detection                  & Yes                                                     & Yes                        & Yes                                          \\ \hline
AI attack detection                & Partial                                                 & No                         & No                                           \\ \hline
Adversarial training               & Yes                                                     & No                         & No                                           \\ \hline
Robustness to adversarial examples & Yes                                                     & No                         & No  \\ \hline
Cyber Threat Intelligence & No                                                     & Yes                         & Yes\\ \hline
Vulnerability analysis & No                                                     & Yes                         & Yes\\ \hline
\end{tabular}
\end{table}

\begin{figure*}
\centering
    \includegraphics[width=0.8\linewidth]{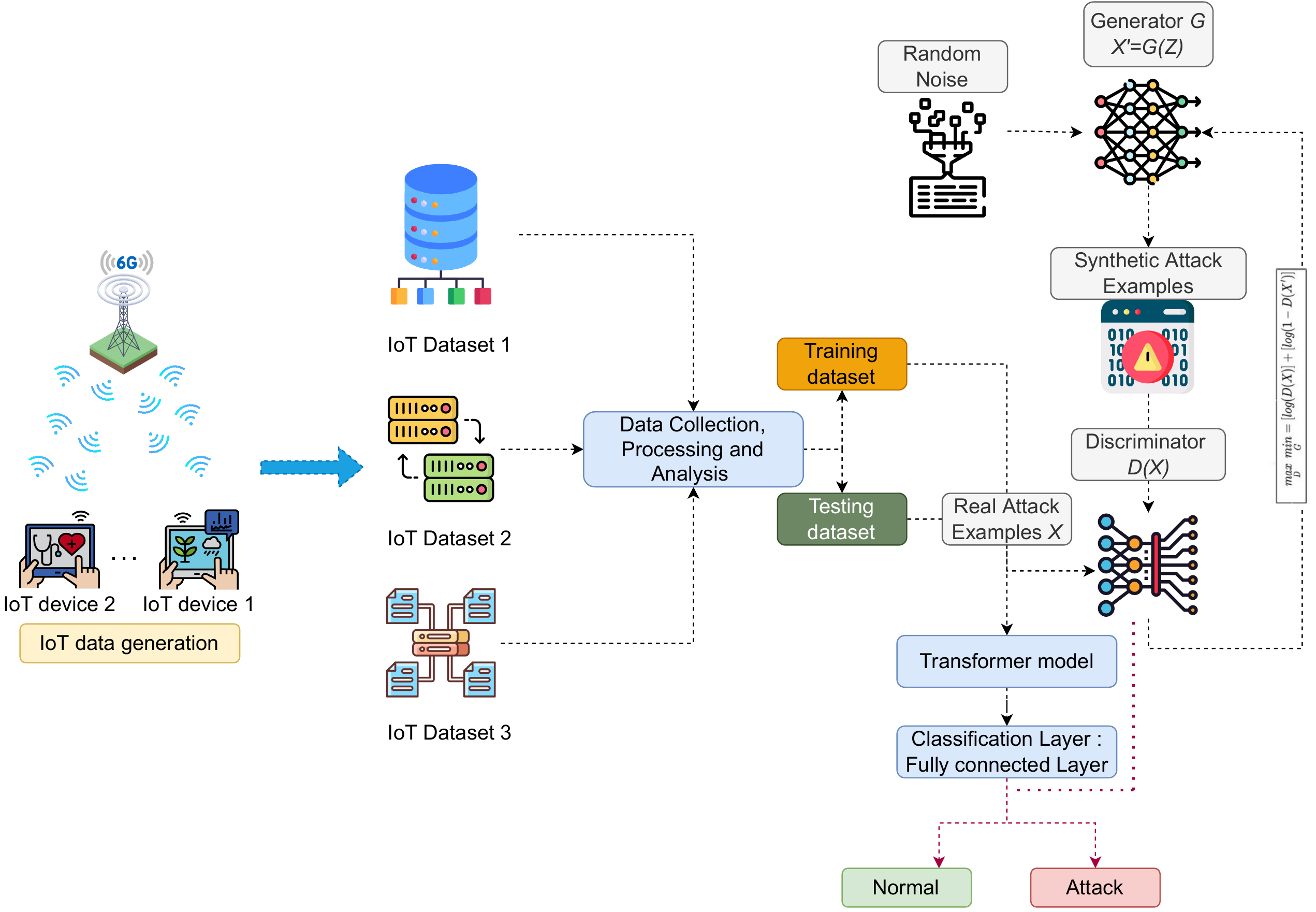}
    \caption{The proposed GAN and Transformer-based model for Cyber Threat-Hunting in 6G-enabled IoT Networks.}
    \label{fig:fig1}
\end{figure*}

\section{Generative AI for Cyber Security}\label{sec:3}

Table \ref{tab:tab1} reviews the recent works on Generative AI for Cyber Threat-Hunting.

\subsection{Generative AI-based method}

\subsubsection{GAN-based method}
Cui et al. \cite{cui2021security} introduced DP-GAN, a modified version of the GAN model that enables decentralized and asynchronous Federated Learning (FL) while preserving differential privacy. The system operates by concurrently running two games: one between the generator and discriminator of a traditional GAN, and the other between the discriminator and a new component named DP identifier (DPI). Additionally, the framework incorporates blockchain technology to enhance system reliability and establish a decentralized IoT anomaly detection system. On the other hand, Block Hunter, developed by Yazdinejad et al. \cite{yazdinejad2022block}, utilizes a cluster-oriented design to identify irregularities in smart factories based on blockchain technology. By employing the cluster-based approach, the detection process becomes more effective as it lowers the amount of data transmitted while also increasing throughput in IIoT networks. Block Hunter also examines various anomaly detection algorithms, including clustering-based, statistical, subspace-based, classifier-based, and tree-based techniques. The proposed approach is evaluated based on block generation, block size, and miners and the assessment includes metrics such as accuracy, precision, recall, F1-score, and True Positive Rate (TPR).

The security challenges of the Internet of Things (IoT) are addressed by Habibi et al. in their study \cite{habibi2023imbalanced}, with a specific focus on the problem of IoT Botnets and their impact on the reliability of IoT systems. The authors assert that current botnet detection methods are flawed as they rely on untrustworthy or unlabeled datasets, leading to a decrease in the performance of security tools. To address this, the study proposes the use of a Generative Adversarial Network model for tabular data modeling and generation. Results indicate that with data augmentation through the proposed Generative AI algorithm, Multilayer Perceptron (MLP) shows high accuracy and F1-score, as well as high sensitivity and specificity. Similarly, the utilization of GAN in trust management for dependable and immediate communication in 6G wireless networks is investigated by Yang et al. \cite{yang2022generative}. A novel intelligent trust management framework that blends fuzzy logic theory and adversarial learning is introduced in the study. Moreover, a trust decision-making model based on GAN is proposed to appraise the credibility of network devices and enhance the robustness of trust management.

Tabassum et al. \cite{tabassum2022fedgan} introduced FEDGAN-IDS, a Federated Deep Learning Intrusion Detection System that utilizes the GAN architecture to identify cyber threats in smart IoT systems. The purpose of FEDGAN-IDS is to resolve the disparity in data distribution for rare classes by using GANs and augmenting local data to improve privacy and performance in distributed training. The authors assessed the effectiveness of FEDGAN-IDS by comparing it to other federated intrusion detection models and conducting experiments on multiple datasets. The outcomes indicate that the proposed model outperforms the latest standalone IDS and reaches convergence more quickly.

\begin{figure}
\centering
    \includegraphics[width=0.9\linewidth]{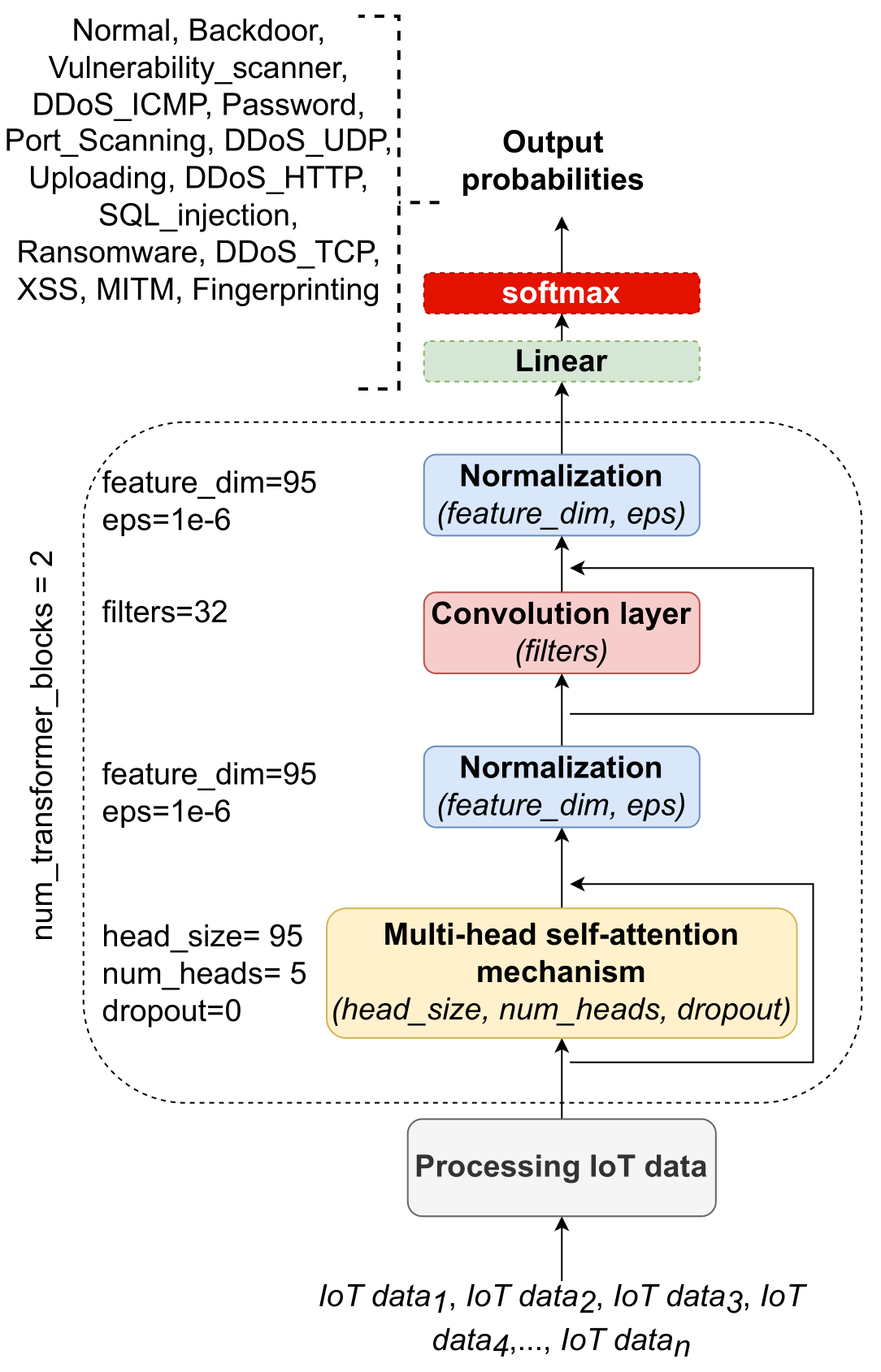}
    \caption{The proposed Transformer - model architecture.}
    \label{fig:fig5}
\end{figure}

Zhang et al. \cite{zhang2020poisongan} conducted a study on the vulnerabilities of federated learning in edge computing for IoT applications. Federated learning is a technique that allows machine learning models to be trained locally instead of centrally to reduce privacy concerns. However, the study revealed that this method is vulnerable to poisoning attacks, where an attacker introduces malicious data to corrupt the global model. To tackle this issue, the study suggested two approaches, namely Data\_Gen and PoisonGAN. Data\_Gen is a technique that utilizes GAN to generate poison data based on the global model parameters. PoisonGAN, on the other hand, is a new poisoning attack model that leverages Data\_Gen to reduce attack assumptions and make the attacks more feasible. The effectiveness of these attack models was tested using two common poisoning attack strategies, label flipping, and backdoor, on a federated learning prototype. The findings demonstrated that these attack models were successful in federated learning.

\subsubsection{GPT-based method}
Ranade et al. \cite{ranade2021generating} investigate the potential harm of incorporating false Cyber Threat Intelligence (CTI) into cyber-defense systems, which rely on semi-structured data and text to create knowledge graphs. They employ GPT-2 with fine-tuning, a type of Transformer, to produce realistic CTI text descriptions and fool the cyber-defense systems. To demonstrate the destructive consequences of this attack, the authors launch a data poisoning assault on a Cybersecurity Knowledge Graph (CKG) and a cybersecurity corpus. The study also involves feedback from cybersecurity professionals and threat hunters, who acknowledge the produced fake CTI as authentic.

\subsubsection{BERT-based method}

Ranade et al. \cite{ranade2021cybert} introduced CyBERT, a specialized version of BERT that has been adapted to the field of cybersecurity. This model utilizes a large corpus of cybersecurity data to improve its performance in processing detailed information related to threats, attacks, and vulnerabilities. The key contribution of this work is the development of a fine-tuned BERT model that can accurately and efficiently complete a range of cybersecurity-specific tasks. The model was trained on open-source, unstructured, and semi-unstructured Cyber Threat Intelligence (CTI) data using Masked Language Modeling (MLM) and was evaluated on various downstream tasks that have potential applications in Security Operations Centers (SOCs). Additionally, the paper also presents examples of how CyBERT can be used in real-world cybersecurity tasks. Jo et al. \cite{jo2022vulcan} introduced Vulcan, a CTI system that extracts relevant information from unstructured text and establishes semantic connections. The system employs neural language model-based named entity recognition and relation extraction techniques. The researchers conducted experiments and found that Vulcan boasts a high level of accuracy, with an F-score of 0.972 for named entity recognition and 0.985 for relation extraction. Furthermore, the system provides a platform for security professionals to create applications for threat analysis, and the study includes two examples of such applications - identifying the evolution of threats and profiling threats - which can save time and resources in analyzing cyber threats and provide in-depth information about the threats.

\subsection{Comparison between Generative AI-based models}
Table \ref{tab:tab2} presents a comparison between Generative AI models, namely, GAN, GPT, and BERT. In order to secure IoT applications, GANs have a number of benefits over GPT and BERT. A key advantage of GANs is their potential to generate new data similar to training data. This leads to their suitability for applications such as phishing detection, spam detection, malware detection, network intrusion detection, and anomaly detection. GANs are also resilient to adversarial examples, making them suitable for attack detection and defense. GPT and BERT, however, are not as robust to adversarial examples as GANs. They are valuable for performing other tasks such as providing natural language understanding and text generation. Currently, proposing a cyber security system using Transformer-based models for securing IoT applications is challenging. For the other cybersecurity metrics listed in the table, GAN, GPT, and BERT lack access control, encryption and decryption capabilities, and authentication. From this analysis, we explore and propose the combination of GAN and Transformer models for Cyber Threat-Hunting in 6G-enabled IoT Networks.

\section{The proposed GAN and Transformer-based model}\label{sec:4}

The proposed GAN and Transformer-based model for Cyber Threat-Hunting in 6G-enabled IoT Networks is presented in Figure  \ref{fig:fig1}. 

\subsection{Generative Adversarial Networks}
Generative Adversarial Networks (GANs) are a type of deep learning algorithm that utilizes generative and discriminative models in combination to generate new data that is similar to an existing dataset. The fundamental structure of a GAN is comprised of two neural networks: a generator and a discriminator. The generator is responsible for creating new data, while the discriminator is tasked with evaluating the authenticity of the generated data. The objective of the generator is to minimize the loss function by generating a greater number of samples that the discriminator classifies as genuine. On the other hand, the discriminator aims to maximize the loss function by accurately identifying as many true data samples as possible and as many generated samples as false.

The GAN algorithm with an IoT cybersecurity dataset can be organized into the following steps  \cite{goodfellow2020generative}:

\begin{itemize}
    \item Step 1: Generator and discriminator networks are initialized: 
     \begin{equation}
         G = G(z) , D = D(x)
     \end{equation} 
    \item Step 2: Generate a random noise vector $z$ from a noise distribution.
    \item Step 3: Pass the noise vector over the generator network to generate a new data point $x'$.
    \item Step 4: Pass the generated data point $x'$ and an actual data point $x$ from the IoT cybersecurity dataset through the discriminator network. 
    \item Step 5: the loss functions of the generator and discriminator networks are calculated as follows: The generator's loss is determined by the negative logarithm of the discriminator's output on the generated sample (G(z)). On the other hand, the discriminator's loss is determined by the negative logarithm of its output on the real sample (x) and the output on the generated sample (G(z)) subtracted from 1.
   \begin{equation}
         L(D) = - (log(D(x)) + log(1-D(G(z))))
     \end{equation}
    \begin{equation}
         L(G) = - log(D(G(z)))
     \end{equation}
    \item Step 6: Update the weights of generator and discriminator networks via backpropagation and optimization techniques, such as gradient descent or Adam's optimization.
    \item Step 7: Repeat steps 2-6 with a fixed iteration or until the loss function achieves a suitable level.
    \item Step 8: Employ the trained generator network to produce more data points that resemble the existing IoT cybersecurity dataset.
\end{itemize}

The objective of the GAN is to reduce the gap between the generated and trained IoT data while increasing the gap between the generated IoT data and the actual IoT data.

\subsection{Transformer model}
Generative Pre-training Transformer (GPT) is a type of Transformer-based neural network language model that is trained using a large dataset of text. It is typically can be used for vulnerability analysis of IoT text data. The algorithm used in GPT for  vulnerability analysis of IoT text data employs the Transformer architecture, which was introduced in a 2017 paper by Google researchers titled "Attention Is All You Need" \cite{vaswani2017attention}. The Transformer architecture is a type of neural network that uses self-attention techniques to process sequence IoT data. Specifically, we consider a set of IoT data $X = {x_1, x_2, ..., x_n}$ as input and a set of labels $Y = {y_1, y_2, ..., y_n}$ as output indicating whether each IoT data attack or not.

The GPT algorithm with an IoT cybersecurity dataset can be organized into the following steps:

\begin{itemize}
    \item Step 1: Data preprocessing. This involves the elimination of any unnecessary text or symbols from IoT data. Then, tokenize the IoT data into individual words as well as eliminate the stop words (e.g.,   "is", "and", "the"). After that, text normalization techniques are adopted (e.g., Lemmatization/Stemming) for further processing.
    \item Step 2: Feature Extraction. This step involves the extraction of features from the preprocessed dataset and represents each IoT data as a feature vector, $X = {x_1, x_2, ..., x_n}$.
    \item Step 3. Fine-tune the GPT model. This step consists of using the train set to fine-tune the GPT model. The model learns the patterns and features of IoT data.
    \item Step 4. Model testing. Once the model is fine-tuned, use the test set to evaluate its performance and classify new IoT data as an attack or not. The model predicts the label for each IoT data, $Y = {y_1, y_2, ..., y_n}$.
    \item Step 5: Evaluation. The model's performance is assessed through metrics such as accuracy, precision, recall, and F1 score. Subsequently, the model's settings are adjusted, and the feature extraction process is modified as necessary to enhance its performance.
\end{itemize}

The architecture of the proposed Transformer model is presented in Figure \ref{fig:fig5}. The architecture consists of two main components: the Transformer Encoder and the Transformer Model. The Transformer Encoder is a module that implements the attention mechanism and feed-forward neural network of a Transformer. It has four main sub-modules: a Layer Norm module for normalizing the input, a Multi-head Attention module for performing self-attention, a Dropout module for regularization, and a Conv1d module for implementing the feed-forward neural network. The architecture uses 95 as the feature dimension and 15 as the number of classes. The Transformer Encoder module is parameterized with head\_size, num\_heads, filters, and dropout. The Transformer Model is parameterized with head\_size, num\_heads, filters, num\_Transformer\_blocks, and dropout.

The attention mechanism adopted by the proposed Transformer model weighs the importance of different elements in the input IoT data, which is defined as:

\begin{equation}
Att(Q, K, V) = softmax(\frac{QK^T}{\sqrt{d_k}})V
\end{equation}

The equation above represents the Attention function, which takes in the query matrix (Q), key matrix (K), and value matrix (V). The output of the function is the dot product of the softmax function of the quotient of the dot product of Q and the transpose of K, divided by the square root of the dimension of the keys $(d_k)$, and the value matrix (V). By calculating the dot product between the query and key matrices, the attention mechanism obtains the attention weights, which are then subjected to the softmax function. To arrive at the weighted sum of the value matrix, the attention weights are employed.

The proposed Transformer model architecture also uses multi-head attention, which is defined as:

\begin{equation}
MultiHead(Q, K, V) = Concat(h_1, ..., h_h)W^O
\end{equation}

Where $h_i$ is computed as:

\begin{equation}
h_i = Att(QW_i^Q, KW_i^K, VW_i^V)
\end{equation}

The formula computes the attention score for the i-th element in a set of query vectors, using the corresponding key and value vectors. The attention score is calculated by feeding the query vector through a linear transformation specified by the matrices Q and K, followed by a softmax operation over the dot products between the transformed query and key vectors. The resulting attention weights are then used to compute a weighted sum of the value vectors, yielding the output vector $h_i$.

The position-wise feed-forward network is defined as:

\begin{equation}
FFN(x) = max(0, xW_1 + b_1)W_2 + b_2
\end{equation}

Where $x$ is the input, $W_1$ and $W_2$ are weight matrices, and $b_1$ and $b_2$ are biases.

The proposed Transformer includes a residual connection and layer normalization. The residual connection is defined as the sum of the input x and the output of the sub-layer. Layer normalization is defined by subtracting the mean $\mu$ from the input x and dividing by the standard deviation $\sigma$ . The residual connection is defined as:

\begin{equation}
residual(x)= x + Sublayer(x)
\end{equation}

Where $x$ is the input and $Sublayer$ is the sub-layer of the Transformer. The layer normalization is defined as:

\begin{equation}
LayerNorm(x) = \frac{x - \mu}{\sigma}
\end{equation}

\section{Experimental Evaluation}\label{sec:5}
The experimental evaluation of generative AI for cyber threat-hunting in 6G-enabled IoT networks is a critical step in understanding the capabilities and limitations of this technology. In this section, we will evaluate the performance of the proposed GAN and Transformer-based model for Cyber Threat-Hunting.

\subsection{Experimental setup and pre-processing of the Dataset}

The Edge-IIoT dataset, sourced from \cite{ferrag2022edge}\footnote{https://www.kaggle.com/datasets/mohamedamineferrag/edgeiiotset-cyber-security-dataset-of-iot-iiot}, is our reference in this study. It is composed of 15 classes, comprising 1 Normal class and 14 attack classes. The dataset was generated from a testbed intended for IoT and IIoT applications, covering diverse devices, sensors, protocols, cloud and edge configurations. The data was sourced from over 10 types of IoT devices, including those for flame detection, heart rate monitoring, soil moisture measurement, pH level tracking, water level detection, ultrasonic detection, and environmental measurement sensors. The dataset further provides 14 types of attacks against IoT and IIoT connectivity protocols, categorized into five categories, namely Denial of Service/Distributed Denial of Service, information gathering, man-in-the-middle attacks, injection attacks, and malware attacks. The dataset also presents 61 diverse features acquired from different sources such as network traffic, logs, system resources, and alerts.

The initial step is to organize and clean the data by eliminating duplicate entries and filling in any missing values. Next, we eliminate any irrelevant features and convert categorical variables using label encoding. After that, we use a standardization technique to normalize the features and divide the data into training and testing sets, with the training set utilized for model validation and the test set reserved for the final evaluation of the model.

\begin{table}[h!]
\setlength{\tabcolsep}{2.5pt}
\renewcommand{\arraystretch}{1}
\caption{Classification report for multi-classification.}
\centering
\begin{tabular}{|c|c|c|c|c|}
\hline
\textbf{Class}  & \multicolumn{1}{c|}{\textbf{Precision}} & \multicolumn{1}{c|}{\textbf{Recall}} & \textbf{F1-Score} & \textbf{Support} \\ \hline\hline
Normal & 1.00 & 1.00 & 1.00 & 323129 \\ \hline
Backdoor & 0.95 & 0.94 & 0.95 & 4972 \\ \hline
Vulnerability\_scanner & 0.94 & 0.85 & 0.89 & 10022 \\ \hline
DDoS\_ICMP & 1.00 & 1.00 & 1.00 & 23287 \\ \hline
Password & 0.43 & 0.80 & 0.56 & 10031 \\ \hline
Port\_Scanning & 0.54 & 0.12 & 0.20 & 4513 \\ \hline
DDoS\_UDP & 1.00 & 1.00 & 1.00 & 22007 \\ \hline
Uploading & 0.61 & 0.38 & 0.47 & 7527 \\ \hline
DDoS\_HTTP & 0.75 & 0.94 & 0.83 & 9982 \\ \hline
SQL\_injection & 0.52 & 0.23 & 0.32 & 10241 \\ \hline
Ransomware & 1.00 & 0.78 & 0.88 & 2185 \\ \hline
DDoS\_TCP & 0.71 & 1.00 & 0.83 & 10012 \\ \hline
XSS & 0.53 & 0.28 & 0.37 & 3183 \\ \hline
MITM & 1.00 & 1.00 & 1.00 & 80 \\ \hline
Fingerprinting & 0.65 & 0.37 & 0.47 & 200 \\ \hline \hline
\textbf{accuracy} & & & \textbf{0.95} & 441371 \\ \hline
\textbf{macro avg} & 0.77 & 0.71& \textbf{0.72} & 441371 \\ \hline
\textbf{weighted avg} & 0.95 & 0.95 & \textbf{0.94} & 441371 \\ \hline
\end{tabular}
\label{tab:tabres1}
\end{table}

\begin{figure}[t!]
\centering
    \includegraphics[width=1\linewidth]{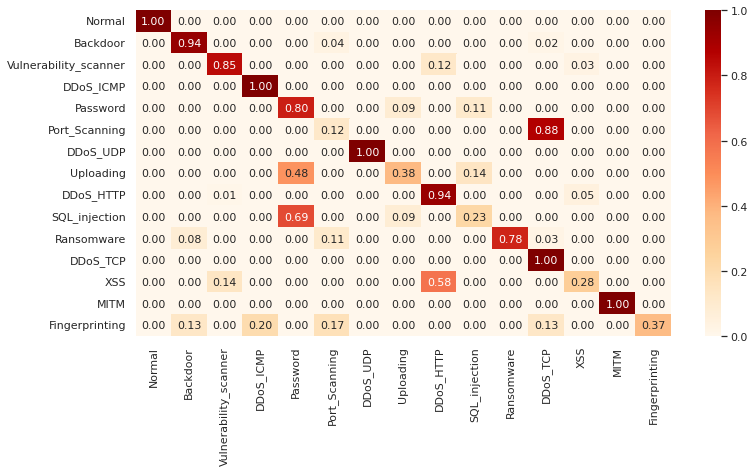}
    \caption{Confusion Matrix for multi-classification.}
    \label{fig:fig3}
\end{figure}

\begin{figure}[t!]
\centering
    \includegraphics[width=1\linewidth]{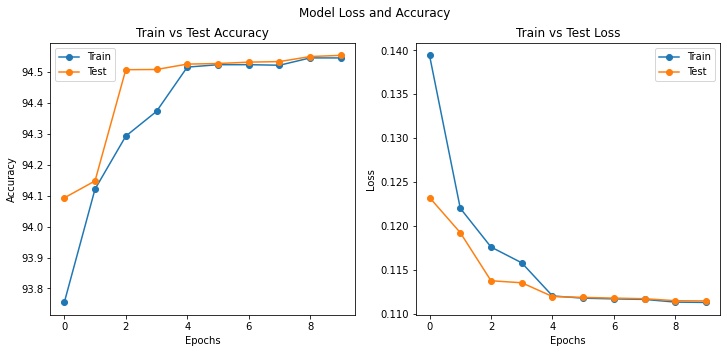}
    \caption{Accuracy and loss for multi-classification.}
    \label{fig:fig4}
\end{figure}

\subsection{Performance Metrics}
In order to evaluate the effectiveness of the proposed model based on GAN and Transformer, the following significant performance metrics are employed:

\begin{itemize}
\item True Positive (TP) refers to the correct identification of attack samples.
\item False Negative (FN) pertains to the incorrect identification of attack samples.
\item True Negative (TN) signifies the accurate identification of benign samples.
\item False Positive (FP) represents the erroneous identification of benign samples.
\item Accuracy measures the ratio of accurately classified entries to the total number of entries, as determined by the formula: 
\begin{equation}
\frac{TP_{Attack}+TN_{Normal}}{TP_{Attack}+TN_{Normal}+FP_{Normal}+FN_{Attack}}
\end{equation}
\item Precision indicates the ratio of correctly classified attack samples to the total number of predicted attack samples, which is determined by the equation:
\begin{equation}
\frac{TP_{Attack}}{TP_{Attack} + FP_{Normal}}
\end{equation}
\item Recall reflects the proportion of accurately identified attack samples to the total number of actual attack samples, as given by the formula:
\begin{equation}
\frac{TP_{Attack}}{TP_{Attack} + FN_{Attack}}
\end{equation}
\item F1-Score represents the harmonic mean of Precision and Recall, as computed by the formula:
\begin{equation}
2 \cdot \frac{Precision \cdot Recall}{Precision+Recall}
\end{equation}
\end{itemize}
\subsection{Experimental Results}

Figure \ref{fig:fig4} presents the evaluation of the model's performance using loss and accuracy metrics. The training loss starts at 0.139 and decreases over time, reaching 0.111 in the last epoch. The training accuracy starts at 93.755\% and increases over time, reaching 94.546\% in the last epoch. This suggests that the model is learning and improving its performance on the training dataset. The testing loss starts at 0.123 and decreases over time, reaching 0.111 in the last epoch. The testing accuracy starts at 94.093\% and increases over time, reaching 94.555\% in the last epoch. This suggests that the model is generalizing well and performing similarly on unseen data. Overall, the model performed well and reached a stable performance of around 95\% accuracy after the 7th Epoch. It's worth noting that the performance on the test dataset is similar to the performance on the training dataset, which means that the model is not overfitting.

Table \ref{tab:tabres1} presents the multi-classification report of the proposed Transformer model for cyber threat-hunting in 6G-enabled IoT networks, which shows a high overall accuracy of 0.95. This indicates that the model is able to accurately identify and classify different types of cyber threats in these networks. The precision, recall, and f1-score of the model also show promising results. The precision metric calculates the percentage of accurate positive predictions made out of all positive predictions made, whereas recall determines the proportion of true positive predictions out of all actual positive occurrences. The f1-score, which is a commonly used indicator of a model's effectiveness, is a harmonic average of precision and recall.

The model shows near-perfect performance in identifying normal instances, with precision, recall, and f1-score of 1.00. It also shows high performance in identifying DDoS\_ICMP and DDoS\_UDP instances, with precision, recall, and f1-score of 1.00. However, the model has lower performance in identifying other types of cyber threats, such as Password, Port\_Scanning, and SQL\_injection. The precision, recall, and f1-score for these types of instances are 0.43, 0.54, and 0.52, respectively. The support metric shows the number of instances for each class, indicating the imbalance in the dataset.

The Confusion Matrix for the proposed Transformer model for multi-classification is shown in Figure \ref{fig:fig3}. We can observe that the normal class pattern was clearly differentiated from all the attack patterns, indicating that the IoT devices' task-oriented nature and consistent data distribution could enhance real-time attack detection capability. Overall, the proposed Transformer model shows promising performance in identifying and classifying cyber threats in 6G-enabled IoT networks. However, there is room for improvement in identifying certain types of cyber threats, such as Password and Port\_Scanning. To improve the performance of the model, additional data and techniques can be used to balance the dataset and further fine-tune the model.

\section{Open Challenges}\label{sec:6}

There are several challenges regarding the use of generative AI for cyber threat-hunting in 6G-enabled IoT networks, including Scalability issues, Decentralized training issues, Data quality issues, Energy challenges, Privacy-preserving challenges, and Tokenization challenges. 

\subsection{Scalability issues}
Generative AI for IoT applications has several scalability issues, including, Cost, Latency, Memory limitations, and High computational requirements. Generative AI is extremely computationally intensive and can be expensive to obtain and maintain. Generative AI models require significant computational resources to train and run, for example, GPT-3 and GPT-4 are large models with 175 billion parameters and 100 trillion parameters, respectively. In addition, the large size of the model also ensures that it requires a lot of memory to run, making it difficult to deploy on memory-limited devices. Therefore, the question we ask here is: how to optimize the scalability of a Generative AI-based system for IoT? We believe that a comparative study of the scalability of Generative AI is needed for IoT security.

\subsection{Decentralized training issues} 
Decentralized training of Generative AI raises concerns about data privacy and security, as IoT data (i.e., sensitive and personal information) may be exposed to malicious nodes. Therefore, the coordination of decentralized Generative AI model training among multiple parties can be complex and time-consuming. One potential area of research in this topic could be focused on creating secure and confidential solutions for AI-generated models within decentralized settings.

\subsection{Data quality issues}
One of the most significant data quality issues when using Generative AI models is data bias. GPT-3 for example is trained on a massive dataset of internet text, which can introduce bias into the model. An additional problem with data quality that can occur when using GPT-3 is data noise. With the large amount of data used to train GPT-3, it is probable that there will be some noise in the data that will be of poor quality or irrelevant to that task. Reducing the impact of this noise on the model's performance is crucial, as it may result in the model learning incorrect patterns or making inaccurate predictions. Consequently, there is a need to prioritize the challenge of mitigating this issue by providing a high-quality and well-organized dataset.

\subsection{Energy challenges}
One of the main challenges of Generative AI models is their computational power. For example, GPT-3 model has 175 billion parameters, which makes it one of the largest language models in existence. To build such a model, a large quantity of computing power is needed. An additional energy challenge of GPT-3 is its implementation. GPT-3 requires a large quantity of memory to operate, which means that it needs to be deployed on high-performance servers. At the same time, these servers use a large amount of power to run, which makes a contribution to the overall energy footprint of the model. To reduce the model's energy consumption during training and deployment, there are some solutions that can be adopted in the future, including, reducing the model's size, adopting federated learning, and employing more energy-efficient hardware (e.g., GPUs or TPUs).

\subsection{Privacy-preserving challenges}
There are various challenges with privacy preservation that are associated with GPT-3, including model extraction, model inversion, and data leakage. GPT-3 can store vulnerable data from fine-tuning data. Attackers can use the text generated by GPT-3 to infer private information from the fine-tuning data or the data used to train the model. The GPT-3 model's parameters can be extracted by attackers and used to generate text or infer privacy from the model. The most important question that may arise is how to develop a new privacy strategy such as Differential privacy.

\subsection{Tokenization challenges}
Transformer-based models are built to process different length input sequences, which makes them adequate for a wide range of Natural language processing (NLP) tasks. The input sequences, however, must be in a suitable format for the model to be processed. That is where Tokenization becomes important, which is the process of splitting the text into single words or sub-words. The main challenges in tokenization for Transformer-based models are dealing with special characters (e.g., punctuation, emoji, and numbers), out-of-vocabulary (OOV) words, and rare words. OOV words are those that are not present in the model's vocabulary, which can cause issues for the model, as it is not able to process them properly. Rare words are those that appear rarely in the training data, which can cause issues for the model to handle them correctly. One potential avenue of research regarding this subject may involve the process of tokenization for IoT traffic.

\section{Conclusions}\label{sec:7}
In this paper, we discussed the use of generative AI for cyber threat-hunting in 6G-enabled IoT networks, which has the potential to revolutionize the way we detect and prevent cyber-attacks. Then, we proposed a new generative adversarial network (GAN) and Transformer-based model for Cyber Threat-Hunting in 6G-enabled IoT Networks. The experimental analysis results with a new cyber security dataset demonstrate that the Transformer-based security model for cyber threat-hunting can detect IoT attacks with a high overall accuracy of 95\%. However, there are also several challenges that must be addressed, including scalability, decentralized training, data quality, energy efficiency, privacy, and tokenization. Despite these challenges, the potential benefits of using generative AI for cyber threat-hunting in 6G-enabled IoT networks make it a promising area of research that is worth exploring further.


\bibliographystyle{IEEEtran}
\bibliography{ref} 

\begin{thebibliography}{10}
\providecommand{\url}[1]{#1}
\csname url@samestyle\endcsname
\providecommand{\newblock}{\relax}
\providecommand{\bibinfo}[2]{#2}
\providecommand{\BIBentrySTDinterwordspacing}{\spaceskip=0pt\relax}
\providecommand{\BIBentryALTinterwordstretchfactor}{4}
\providecommand{\BIBentryALTinterwordspacing}{\spaceskip=\fontdimen2\font plus
\BIBentryALTinterwordstretchfactor\fontdimen3\font minus
  \fontdimen4\font\relax}
\providecommand{\BIBforeignlanguage}[2]{{%
\expandafter\ifx\csname l@#1\endcsname\relax
\typeout{** WARNING: IEEEtran.bst: No hyphenation pattern has been}%
\typeout{** loaded for the language `#1'. Using the pattern for}%
\typeout{** the default language instead.}%
\else
\language=\csname l@#1\endcsname
\fi
#2}}
\providecommand{\BIBdecl}{\relax}
\BIBdecl

\bibitem{gerodimos2023iot}
A.~Gerodimos, L.~Maglaras, M.~A. Ferrag, N.~Ayres, and I.~Kantzavelou, ``Iot:
  Communication protocols and security threats,'' \emph{Internet of Things and
  Cyber-Physical Systems}, 2023.

\bibitem{babaghayou2021whisper}
M.~Babaghayou, N.~Labraoui, A.~A. Abba~Ari, M.~A. Ferrag, L.~Maglaras, and
  H.~Janicke, ``Whisper: A location privacy-preserving scheme using
  transmission range changing for internet of vehicles,'' \emph{Sensors},
  vol.~21, no.~7, p. 2443, 2021.

\bibitem{dhariwal2020jukebox}
P.~Dhariwal, H.~Jun, C.~Payne, J.~W. Kim, A.~Radford, and I.~Sutskever,
  ``Jukebox: A generative model for music,'' \emph{arXiv preprint
  arXiv:2005.00341}, 2020.

\bibitem{zhang2020poisongan}
J.~Zhang, B.~Chen, X.~Cheng, H.~T.~T. Binh, and S.~Yu, ``Poisongan: Generative
  poisoning attacks against federated learning in edge computing systems,''
  \emph{IEEE Internet of Things Journal}, vol.~8, no.~5, pp. 3310--3322, 2020.

\bibitem{ranade2021cybert}
P.~Ranade, A.~Piplai, A.~Joshi, and T.~Finin, ``Cybert: Contextualized
  embeddings for the cybersecurity domain,'' in \emph{2021 IEEE International
  Conference on Big Data (Big Data)}.\hskip 1em plus 0.5em minus 0.4em\relax
  IEEE, 2021, pp. 3334--3342.

\bibitem{wu2021intelligent}
Y.~Wu, L.~Nie, S.~Wang, Z.~Ning, and S.~Li, ``Intelligent intrusion detection
  for internet of things security: A deep convolutional generative adversarial
  network-enabled approach,'' \emph{IEEE Internet of Things Journal}, 2021.

\bibitem{cui2021security}
L.~Cui, Y.~Qu, G.~Xie, D.~Zeng, R.~Li, S.~Shen, and S.~Yu, ``Security and
  privacy-enhanced federated learning for anomaly detection in iot
  infrastructures,'' \emph{IEEE Transactions on Industrial Informatics},
  vol.~18, no.~5, pp. 3492--3500, 2022.

\bibitem{ranade2021generating}
P.~Ranade, A.~Piplai, S.~Mittal, A.~Joshi, and T.~Finin, ``Generating fake
  cyber threat intelligence using transformer-based models,'' in \emph{2021
  International Joint Conference on Neural Networks (IJCNN)}.\hskip 1em plus
  0.5em minus 0.4em\relax IEEE, 2021, pp. 1--9.

\bibitem{yang2022generative}
L.~Yang, Y.~Li, S.~X. Yang, Y.~Lu, T.~Guo, and K.~Yu, ``Generative adversarial
  learning for intelligent trust management in 6g wireless networks,''
  \emph{IEEE Network}, vol.~36, no.~4, pp. 134--140, 2022.

\bibitem{tabassum2022fedgan}
A.~Tabassum, A.~Erbad, W.~Lebda, A.~Mohamed, and M.~Guizani, ``Fedgan-ids:
  Privacy-preserving ids using gan and federated learning,'' \emph{Computer
  Communications}, vol. 192, pp. 299--310, 2022.

\bibitem{jo2022vulcan}
H.~Jo, Y.~Lee, and S.~Shin, ``Vulcan: Automatic extraction and analysis of
  cyber threat intelligence from unstructured text,'' \emph{Computers \&
  Security}, p. 102763, 2022.

\bibitem{habibi2023imbalanced}
O.~Habibi, M.~Chemmakha, and M.~Lazaar, ``Imbalanced tabular data modelization
  using ctgan and machine learning to improve iot botnet attacks detection,''
  \emph{Engineering Applications of Artificial Intelligence}, vol. 118, p.
  105669, 2023.

\bibitem{friha2021internet}
O.~Friha, M.~A. Ferrag, L.~Shu, L.~Maglaras, and X.~Wang, ``Internet of things
  for the future of smart agriculture: A comprehensive survey of emerging
  technologies,'' \emph{IEEE/CAA Journal of Automatica Sinica}, vol.~8, no.~4,
  pp. 718--752, 2021.

\bibitem{yazdinejad2022block}
A.~Yazdinejad, A.~Dehghantanha, R.~M. Parizi, M.~Hammoudeh, H.~Karimipour, and
  G.~Srivastava, ``Block hunter: Federated learning for cyber threat hunting in
  blockchain-based iiot networks,'' \emph{IEEE Transactions on Industrial
  Informatics}, vol.~18, no.~11, pp. 8356--8366, 2022.

\bibitem{goodfellow2020generative}
I.~Goodfellow, J.~Pouget-Abadie, M.~Mirza, B.~Xu, D.~Warde-Farley, S.~Ozair,
  A.~Courville, and Y.~Bengio, ``Generative adversarial networks,''
  \emph{Communications of the ACM}, vol.~63, no.~11, pp. 139--144, 2020.

\bibitem{vaswani2017attention}
A.~Vaswani, N.~Shazeer, N.~Parmar, J.~Uszkoreit, L.~Jones, A.~N. Gomez,
  {\L}.~Kaiser, and I.~Polosukhin, ``Attention is all you need,''
  \emph{Advances in neural information processing systems}, vol.~30, 2017.

\bibitem{ferrag2022edge}
M.~A. Ferrag, O.~Friha, D.~Hamouda, L.~Maglaras, and H.~Janicke,
  ``Edge-iiotset: A new comprehensive realistic cyber security dataset of iot
  and iiot applications for centralized and federated learning,'' \emph{IEEE
  Access}, vol.~10, pp. 40\,281--40\,306, 2022.

\end{thebibliography}

\end{document}